\documentclass[prb,showpacs,aps,10pt]{revtex4}
\usepackage[dvips]{graphicx}
\usepackage{epsfig}
\usepackage{color}
\usepackage{amsmath,amssymb}
\usepackage{bm}
\usepackage[next]{inputenc}

\def\be{\begin{equation}}
\def\ee{\end{equation}}

\def\bi{\begin{itemize}}
\def\ei{\end{itemize}}
\def\bn{\begin{enumerate}}
\def\en{\end{enumerate}}
\def\bea{\begin{eqnarray}}
\def\eea{\end{eqnarray}}

\def\ba{\begin{array}}
\def\ea{\end{array}}
\def\bd{\begin{displaymath}}
\def\ed{\end{displaymath}}

\begin{document}
\title{Phase diagram of the XXZ ferrimagnetic spin-(1/2, 1) chain in the presence of transverse
magnetic field}

\author{A. Langari$^1$, J. Abouie$^{2,3}$, M. Z. Asadzadeh$^1$ and M. Rezai$^1$}
\address{$^1$ Department of Physics, Sharif University of Technology,
Tehran 11155-9161, Iran}
\address{$^2$ Department of Physics, Shahrood University of Technology, Shahrood 36199-95161,
Iran}
\address{$^3$ School of Physics, Institute for Research in Fundamental Sciences (IPM), 
Tehran 19395-5531, Iran}

\begin{abstract}
We investigate the phase diagram of an anisotropic ferrimagnetic spin-($1/2, 1$)
in the presence 
of a non-commuting (transverse) magnetic field. We find a magnetization plateau
for the isotropic case while there is no plateau for the anisotropic ferrimagnet. 
The magnetization plateau can appear only when the Hamiltonian has the
U(1) symmetry in the presence of the magnetic field.
The anisotropic model is driven by the magnetic field from 
the N\'{e}el phase for low fields to the spin-flop phase for 
intermediate fields and then to the paramagnetic phase for high fields.
We find the quantum critical points and their dependence on the anisotropy of the 
aforementioned field-induced quantum phase transitions.
The spin-flop phase corresponds to the spontaneous breaking of Z$_2$ symmetry.
We use the numerical
density matrix renormalization group and analytic spin wave theory to find the phase 
diagram of the model. The energy gap, sublattice magnetization, and total magnetization
parallel and perpendicular to the magnetic field are also calculated. 
The elementary excitation
spectrums are obtained via the spin wave theory in the three different regimes 
depending on the strength of the magnetic field.

\end{abstract}
\date{\today}


\maketitle
\section{Introduction \label{introduction}}

Quantum ferrimagnets are a general class of strongly correlated magnetism,
which have attracted  much interest in experimental as well as theoretical investigations.
Examples of such realizations are the bimetallic molecular magnets like
CuMn(S$_2$C$_2$O$_2$)$_2$(H$_2$O)$_3\cdot$4.5H$_2$O and numerous bimetallic chain compounds
which have been synthesized systematically \cite{Gleizes 81,Pei 87}.
In these materials, the unit cell of the magnetic system is composed of two spins,
the smaller one is $\sigma=1/2$ and the larger one ($\rho$) is changed from
$1/2$ to $5/2$. The magnetic and thermodynamic properties of these models are 
different from the
homogeneous spin counterparts. For instance, the one dimensional mixed-spin model represents
a ferromagnetic behavior for the low temperature regime while a crossover appears
to the antiferromagnetic behavior as temperature increases
\cite{Pati 97, Yamamoto 99, Kolezhuk 99, Jahan2004, Jahan2006}.
The crossover can be explained in terms of the two elementary excitations where the lower one
has the ferromagnetic nature and a gapped spectrum above it with antiferromagnetic
property \cite{Yamamoto 98}.
Moreover, the mixed spin models have shown interesting behavior for the quasi one dimensional
lattices (ferrimagnetic ladders). Despite that the two-leg spin-1/2 ladder is gapful,
representing a Haldane phase, the two-leg (mixed spin) ferrimagnet is always gapless with 
the ferromagnetic nature in the low energy spectrum. However, a special kind of dimerization
can drive the ferrimagnetic ladder to a gapped phase \cite{Langari 00l, Langari 01}. 

The presence of a longitudinal magnetic field preserves the U(1) symmetry of the XXZ interactions
and creates a nonzero magnetization plateau in a one-dimensional ferrimagnet for small
magnetic fields in addition to the saturation plateau for large magnetic fields
\cite{Alcaraz 97, Sakai 99,  Abolfath 01}.
The former plateau corresponds to the opening of the Zeeman energy gap which removes the 
high degeneracy of the ground state subspace. 
The ferrimagnets on ladder geometry present a rich structure of plateaus
depending on the ratio and dimerization of exchange couplings \cite{Langari 00p}.
In both one-dimensional and two-leg ferrimagnets the magnetization plateaus can be understood
in terms of the Oshikawa, Yamanaka and Affleck (OYA) argument \cite{OYA}
because the longitudinal magnetic field commutes with the rest of 
the Hamiltonian and the models have U(1) symmetry. However, the situation is different
when a transverse magnetic field is applied on the system, because the transverse field does not
commute with the XXZ interaction and breaks the U(1) symmetry of the model. 
The onset of a transverse field 
develops an energy gap in a spin-1/2 chain which
initiates an antiferromagnetic order perpendicular to the field 
direction \cite{Dmitriev 02, Essler 03, Langari 04, Dmitriev 04}. 
The ordered phase is a spin-flop
phase because of nonzero magnetization in the field direction; however, there is 
no magnetization plateau even in the gapped phase \cite{Langari 06}. 
The lack of U(1) symmetry prohibits
the use of the OYA argument, 
thus prompts the question of a magnetization
plateau and the presence of an energy gap \cite{Oshikawa2000} in the spectrum. 

The structure of the paper is as follows.
First we study the anisotropic ferrimagnetic chain in the presence of a transverse
magnetic field by using the density matrix renormalization group (DMRG) \cite{White1993}
and exact diagonalization Lanczos methods. 
The energy gap, sublattice
magnetization, and total magnetization in both parallel and perpendicular to the field direction
are presented in Sec. \ref{dmrg}. We further address the energy gap behavior versus
the magnetic field and the magnetization plateau. The phase diagram of
the model is also presented in the same section. We then use an analytical tool,
the spin wave theory (SWT), to obtain the low energy excitation spectrum of the model
in Sec. \ref{swt}.  The SWT is applied in three different regions
depending on the strength of the magnetic field. The qualitative behavior of the model
is explained in terms of SWT and the magnetization is compared with DMRG results.
The results of SWT help to explain the energy gap behavior of DMRG data.
We finally summarize our results in Sec. \ref{summary}, where we put together
both quantitative DMRG and
qualitative SWT results to analyze the different phases of the model
in the presence of a transverse magnetic field.


\section{Density Matrix Renormalization Group results\label{dmrg}}

We have implemented the numerical DMRG technique to study
the magnetic properties of the anisotropic ferrimagnetic spin-($1/2, 1$) 
chain in the presence of
a transverse magnetic field given by the Hamiltonian (\ref{hamiltonian}):
\begin{equation}
\label{hamiltonian}
H=J\sum_{i=1}^{N} [ \sigma^x_{i}\rho^x_{i}+\sigma^y_i\rho^y_{i}+\sigma^x_{i}\rho^x_{i+1}+\sigma^y_{i}\rho^y_{i+1}
+\Delta( \sigma^z_i\rho^z_{i}+\sigma^z_{i}\rho^z_{i+1})-h (\sigma^x_i+\rho^x_{i}) ],
\end{equation}
where $\sigma_i^{\alpha}$ ($\rho_i^{\alpha}$) represents the $\alpha$-component of spin operators
at site $i$ for spin amplitude $\sigma=1/2$ ($\rho=1$). The antiferromagnetic
exchange coupling is $J>0$, the anisotropy is defined by $\Delta$, and $h$ is proportional
to the strength of the transverse magnetic field. 

The DMRG computations have been done on an open chain of length $108$ spins ($N=54$ unit cells) and
the number of states kept in each step of DMRG is $300 \leq m \leq 500$.
We have also studied the chains with larger lengths (up to $N=100$) and observed
no significant changes on the data of magnetization and staggered magnetization 
within 5 digits of accuracy.

\begin{figure}
\vspace{1cm}
\begin{center}
\includegraphics[width=8cm,angle=0]{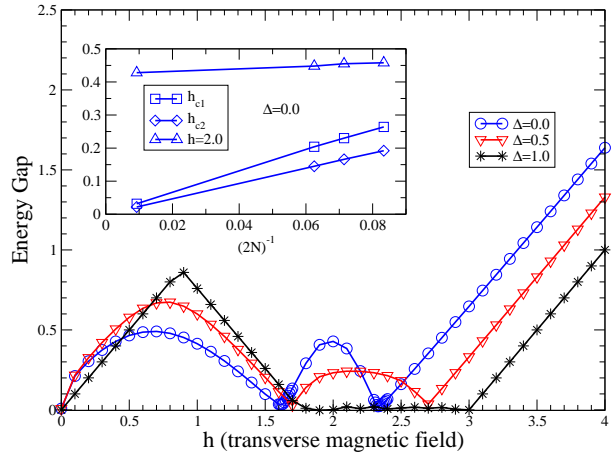}
\caption{The energy gap versus the transverse
magnetic field. Different plots belong to various values of the anisotropy
parameter $\Delta=0.0, 0.5. 1.0$. Inset: The scaling of gap versus $(2N)^{-1}$ for
$\Delta=0.0$ at two critical points $h_{c1}$ and $h_{c2}$ confirms the vanishing
of gap at these points while its scaling at $h=2.0$ verifies a finite gap in the
thermodynamic limit ($N\rightarrow \infty$).
}
\label{gap}
\end{center}
\end{figure}
The energy gap is defined as the difference between the first excited state
energy and the ground state energy. It shows whether the model is gapless
or gapful depending on its zero or nonzero value, respectively.
Using the DMRG computations, we have plotted in Fig. \ref{gap} 
the energy gap of the model versus the transverse magnetic field for different values of anisotropy 
parameter, $\Delta=0, 0.5, 1.0$.
All plots show a gapped phase for small values of the magnetic field, $h < h_{c1}(\Delta)$, 
and a paramagnetic gapped phase for $h >  h_{c2}(\Delta)$. 
The gap vanishes at two critical points, $h = h_{c1}(\Delta)$ and $h = h_{c2}(\Delta)$.
The isotropic case ($\Delta=1$) remains gapless in the intermediate 
region $h_{c1}(\Delta) < h  < h_{c2}(\Delta)$, while the anisotropic case ($\Delta\neq1$)
is gapful.
The gap behaves differently for various $\Delta$
in the small-field and intermediate-field gapped phase. 

In the isotropic case $\Delta=1$, the behavior of gap versus $h$ can be explained in 
terms of the elementary excitations of the model. 
For $\Delta=1$, the U(1) symmetry of the model is restored
and the magnetic field operator commutes with the rest of the Hamiltonian. Thus, the
energy spectrum for $\Delta=1$ is expressed in terms of the spectrum at $h=0$ plus a shift of
energy which depends on $h$.
At $h=0$ the model has SU(2) symmetry and the ground state is a ferromagnetic state
with total spin $S_G=N(|\rho-\sigma|)$, which is highly degenerate and the lowest ferromagnetic spectrum is a 
gapless one, namely $\nu^-(k)$ [see Eq. (\ref{WFSWT})]. 
An antiferromagnetic spectrum ($\nu^+(k)$)  exists above the ferromagnetic one,
and the lowest state of the antiferromagnetic spectrum has total spin $S_{AF}=S_G+1$
with a finite gap $2 J |\rho - \sigma|$, measured from the ground state.
Upon adding a commuting magnetic field to the ferrimagnetic chain
the symmetry is lowered to U(1) and the energy levels are affected
by a Zeemann term, i.e., $ -h S^x$. For the magnetic fields smaller than $\overline{h}$ 
(which will be defined later), the Zeeman energy gain
of the ground state is larger than all of the other states in the ferromagnetic spectrum; thus
the ground state remains robust, and the first excited state is the first state in the
ferromagnetic spectrum (with energy $J(\nu^-(k)+h)$), which leads to the
energy gap equal to $J h$. This explanation remains
valid until the gain of the Zeeman term of the lowest state of the 
antiferromagnetic spectrum ($J(\nu^+(k)-h)$)
dominates the gain of the first excited state in the ferromagnetic spectrum.
It defines $\overline{h}$ by the following equation:
\be
\nu^-(k)+\overline{h}=\nu^+(k)-\overline{h},
\ee
which gives $\overline{h}=|\rho-\sigma|$ within linear approximation of SWT
(from which both $\nu^{\pm}(k)$ will be derived in the next sections).
At this point,
the first excited state is the lowest state of the antiferromagnetic spectrum.
Thus, the energy gap behaves as ($2J|\rho-\sigma| - Jh$) before it vanishes at
 $h=h_{c1}(\Delta=1)=2|\rho-\sigma|$.
The linear increasing behavior for small fields and then linear decreasing of the energy
gap are clear in the DMRG data for $\Delta=1$, shown in  Fig. \ref{gap}. 
Although the DMRG values for
$\overline{h}$ and $h_{c1}$ have some discrepancies with 
those obtained by linear SWT, the SWT gives the
qualitative behavior correctly. 

The energy gap of the anisotropic Hamiltonian ($\Delta\neq1$) is defined as $E_1-E_0$ 
for $0<h<h_{c1}$ and $h>h_{c2}$,
where $E_1$ is the first excited state energy and $E_0$ is the ground state energy.
However, the ground state becomes degenerate ($E1=E_0$) for $h_{c1}\leq h \leq h_{c2}$,
where the energy gap is the difference between the second excited state energy and
the ground state one, $E_2-E_0$.
For small magnetic fields the scaling behavior of the energy gap can be explained 
using the quasi-particle excitations of the model as $h\rightarrow 0$.
The leading term of quasi-particle excitations
for very small magnetic fields ($h \rightarrow 0$) 
gives the scaling of energy gap as $\sqrt{h}$, for $\Delta\neq1$
[in the weak field SWT, Eq.(\ref{weakspectrum})]. 
In a similar manner, the leading term
of the strong field SWT [Eq.(\ref{strongspectrum})] leads to 
linear dependence of the gap on the 
magnetic field in the paramagneic phase which
explains very well the behavior in Fig. \ref{gap}.
The linear dependence of gap versus the magnetic field for 
$h>h_{c2}$ is confirmed by the DMRG numerical data for any isotropies.

\begin{figure}
\vspace{1cm}
\begin{center}
\includegraphics[width=7cm,angle=0]{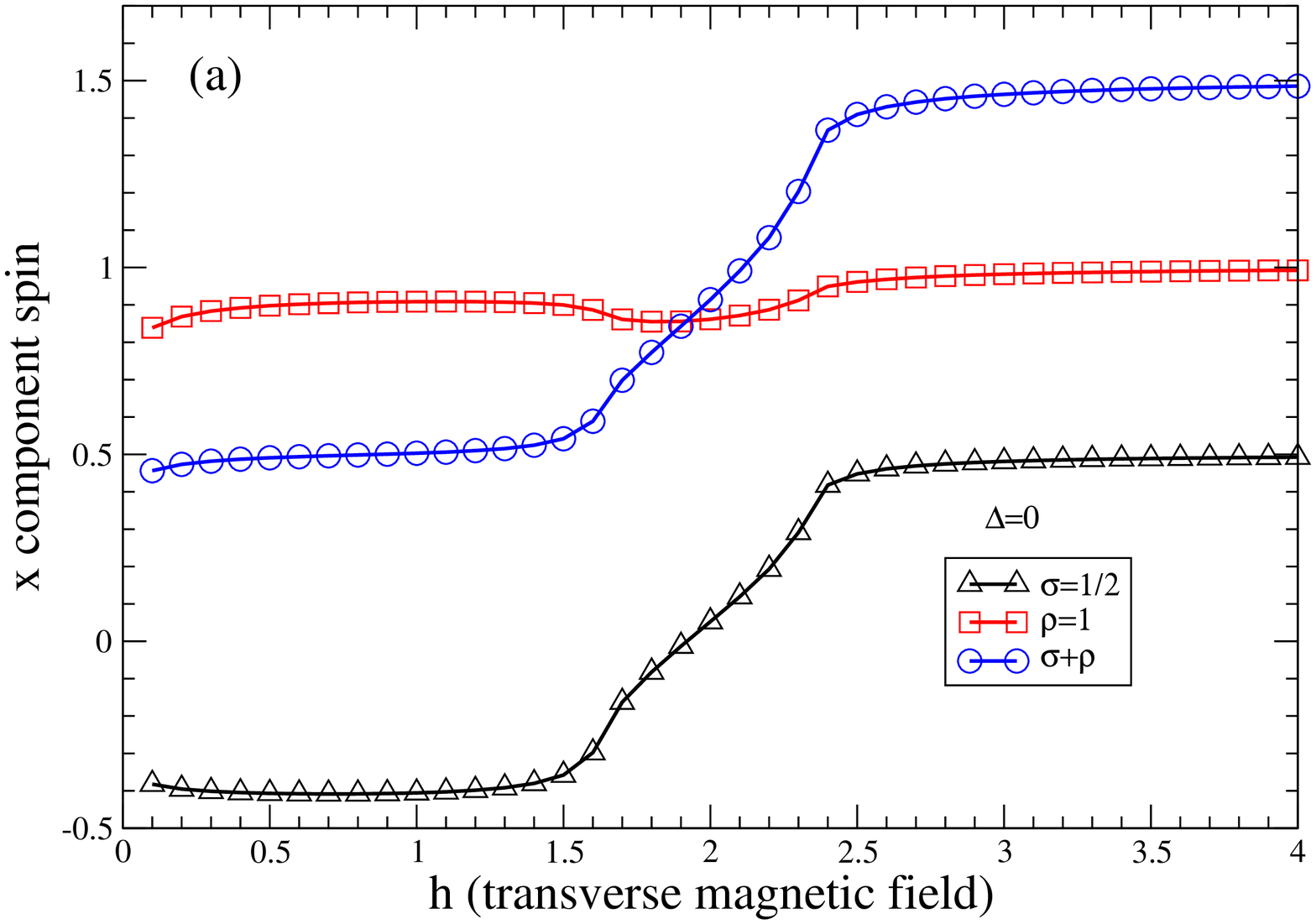}
\includegraphics[width=7cm,angle=0]{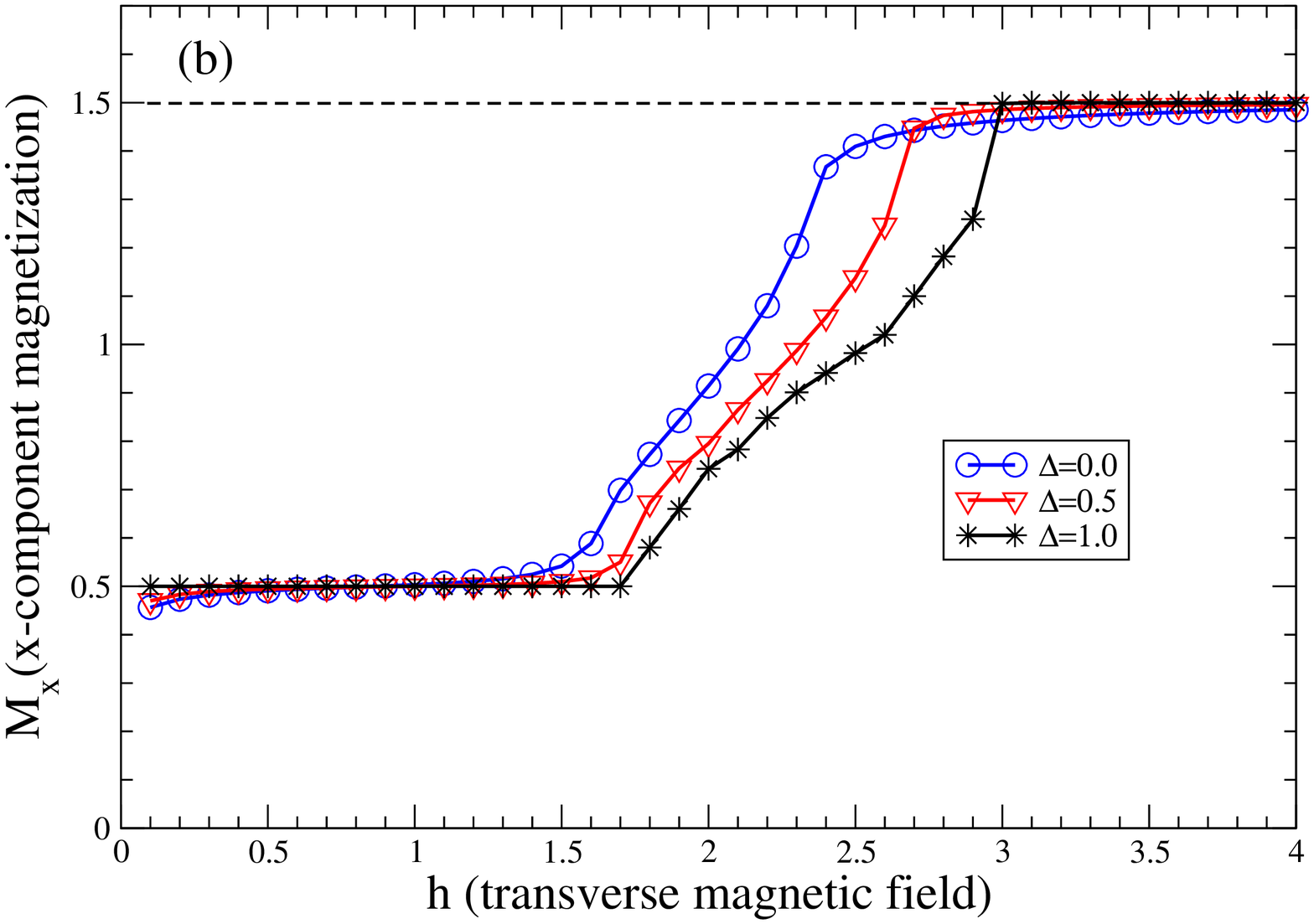}
\caption{(a) The $x$-component sublattice magnetization versus a transverse magnetic field
for both $\sigma=1/2$
and $\rho=1$ spins and their sum as the unit cell magnetization in $x$-direction
for anisotropy parameter $\Delta=0$.
(b) The $x$-component unit cell magnetization versus a transverse magnetic field 
for different anisotropies $\Delta=0.0, 0.5, 1.0$. The dashed line shows the saturation 
value at $M_x=3/2$.
}
\label{xmagnetization}
\end{center}
\end{figure}

We have plotted the energy gap versus $(2N)^{-1}$ in the 
inset of Fig. \ref{gap} to observe its finite size scaling 
(where $2N$ is the total number of spins). We have implemented
both the Lanczos and DMRG algorithms to calculate the energy gap for $\Delta=0$.
We have plotted the minimum value of gap which occurs at $h_{c1}$ and $h_{c2}$ 
versus $(2N)^{-1}$ which clearly shows that the gap vanishes in the thermodynamic limit 
($N \rightarrow \infty$). It suggests that both  $h_{c1}$ and $h_{c2}$ correspond
to quantum critical points. The different magnetization characteristic confirms
that a quantum phase transition occurs at both $h_{c1}$  and $h_{c2}$ 
(see Fig. \ref{xmagnetization}).
We have also plotted the energy gap  for $h=2.0$ to justify that
the gap of the intermediate region is finite in the thermodynamic limit.

We have also plotted the $x$-component magnetization of each sublattice in 
Fig. \ref{xmagnetization}-(a) 
for ferrimagnetic spin-$(1/2, 1)$ chain with $\Delta=0$ versus $h$ employing the DMRG technique. 
The total magnetization has been plotted in Fig. \ref{xmagnetization}-(b)
for different values of anisotropy, $\Delta=0, 0.5, 1.0$.
To calculate the
magnetization we have considered those spins which are far from the open ends
of the chain to avoid the finite size boundary conditions. In this respect, ten spins
have been neglected from each side of the open chain and the
magnetization has been averaged over the rest of spins.
Figure \ref{xmagnetization}-(b) shows the possibility of two plateaus 
in the magnetization along the field direction.
For the isotropic case ($\Delta=1$), it can be explained in terms of 
the OYA argument \cite{OYA}.
According to this argument,  $n(S-m)=\mbox{integer}$, where
$n$ is the periodicity of the ground state, $S$ the total spin of unit cell, 
and $m$ a possible
magnetization plateau of the unit cell, the one-dimensional spin-($1/2, 1$) chain can show two plateaus
at $m=1/2$ and $3/2$. However, for $\Delta\neq 1$ the axial symmetry of the model 
is broken by the transverse magnetic field, and the OYA argument is not 
applicable. Thus, more investigations is required to figure out the difference between
the anisotropic ($\Delta \neq 1$) and isotropic ($\Delta=1$) cases.

\begin{figure}
\vspace{1cm}
\begin{center}
\includegraphics[width=8cm,angle=0]{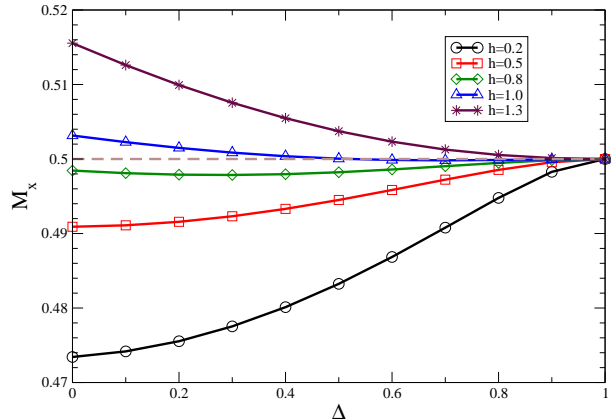}
\caption{Unit cell magnetization ($M_x$) versus the anisotropy parameter ($\Delta$)
for some low magnetic field values ($h$). Our plots justify the plateau only for $\Delta=1$.
The dashed line represents $M_x=0.5$ (the plateau value).
}
\label{xmvsdelta}
\end{center}
\end{figure}

To get more knowledge on the behavior of magnetization
for the anisotropic case, we have plotted the total magnetization in the magnetic 
field direction ($M_x$)
versus the anisotropy parameter ($\Delta$) for small magnetic field values 
in Fig. \ref{xmvsdelta}. The plots have been shown for those values of the magnetic field
which seems to exhibit the magnetization plateaus. Figure \ref{xmvsdelta} clearly verifies that the
magnetization plateau only exists for the isotropic case, while there is no plateau for $\Delta\neq1$.
The magnetization per unit cell ($M_x$) in the direction of magnetic field ($h$)
is given by 
\be
M_x= - \frac{1}{N}\frac{\partial E_0}{\partial h},
\label{mxe0h}
\ee  
where $E_0$ is the ground state energy. 
The above relation for a gapped phase simply states that if the ground state energy is linear in the
magnetic field ($E_0 \propto h$), the magnetization will be constant, (the presence of plateau);
otherwise the magnetization will depend on the magnetic field, (the absence of plateau).
Let write the Hamiltonian as $H=H_0 - h H_1$ 
where $H_0$ is the XXZ interacting part and $h H_1$ is the magnetic field part.
In the presence of U(1) symmetry ($\Delta=1$) the interacting and the magnetic field parts 
commute $[H_0, H_1]=0$. Thus, $E_0$ is a linear function of $h$ which leads to
the emergence of a magnetization plateau when the energy gap is nonzero. 
This agrees with the OYA 
statement. However, the transverse magnetic field breaks the U(1) symmetry in the anisotropic
case ($\Delta\neq1$) and $[H_0, H_1]\neq0$. Therefore, the ground state
energy depends on $h$ non-linearly which gives a change of magnetization
when $h$ varies, i.e. the lack of magnetization plateau even if a finite energy
gap exists.


Although the above general explanation  is applied to the strong
magnetic field regime the saturated plateau ($M_x=1.5$) 
can also be explained from another point of view.
An eigenstate with full saturation is classified as a factorized state \cite{Rezai 10} in which
all spins align in the direction of the magnetic field. 
As a general argument, it has been shown in Ref. \cite{Rezai 10} 
that the full saturation for an anisotropic Heisenberg type interaction in the
presence of a magnetic field takes place at a finite value of the magnetic field
if the model is rotationally invariant around the field direction. Accordingly,
the saturation at $M_x=1.5$ takes place
only for the isotropic case ($\Delta=1$) and $h\geq h_{c2}$. 
In the anisotropic case ($\Delta\neq1$), the fully polarized plateau
can take place for infinite strong magnetic field while the nearly saturated state,
($M_x\simeq 1.5$),  
can be observed for large magnetic fields. 
To justify this argument we have plotted the $x$-component magnetization
of each unit cell for different values of $\Delta$ in Fig. \ref{xmagnetization}-(b).
It is clear that the magnetization in the field direction does not reach the saturation
value of $M_x=1.5$ for $\Delta=0$ and $0.5$, while it obviously touches its saturated value
for $\Delta=1$ and $h\geq3$.

\begin{figure}
\begin{center}
\includegraphics[width=8cm,angle=0]{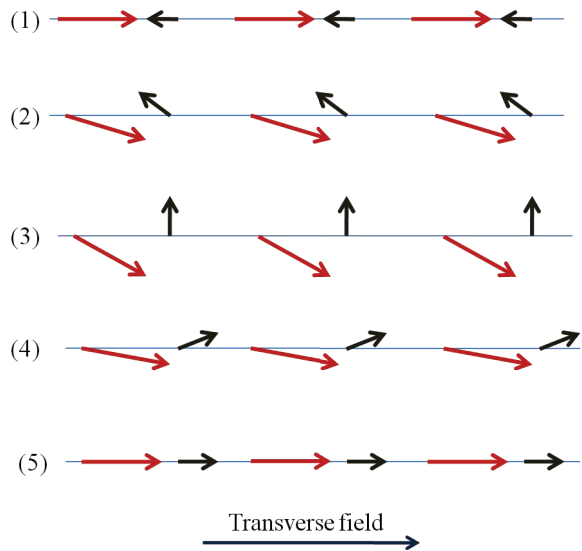}
\caption{Schematic of spins' orientations in different phases of the 
anisotropic ferrimagnetic spin-($1/2,1$) chain in the presence of a transverse magnetic field. 
}
\label{SFS}
\end{center}
\end{figure}

The antiferromagnetic interactions between the spins in each unit cell make them to be antiparallel,
which leads to the total $x$-component magnetization 
$M_x=\langle \sigma_x+\rho_x \rangle \simeq0.5$. 
This phase has been shown schematically in Fig. \ref{SFS}-(1) where we have neglected 
the effects of small quantum fluctuations on the directions of the spins.
The non-commuting transverse magnetic field
opens a gap which is robust as long as $h<h_{c1}$. This (gapped) N\'{e}el phase corresponds
to the first plateau at $M_x=0.5$ for $\Delta=1$ and a semi-plateau ($M_x\simeq0.5$) for 
$\Delta\neq1$.
By further increasing  $h$,  the gap is closed at 
the first critical field $h_{c1}(\Delta)$ (for $\Delta=0$, $h_{c1}\simeq1.6$) 
where the magnetization starts to increase obviously. Further increasing
of the magnetic field leads to a continuous change of the ground state property
which gives a gradual change of the magnetization-Fig. \ref{SFS}-(2-4). 
For strong magnetic field
($h_{c2}(\Delta=0)\gtrsim2.4$) the spins are nearly
aligned in the direction of the magnetic field,
the semi-plateau at $M_x\simeq 1.5$ [Fig. \ref{xmagnetization}-(a) and Fig. \ref{SFS}-(5)].


\begin{figure}
\vspace{1cm}
\begin{center}
\includegraphics[width=8cm,angle=0]{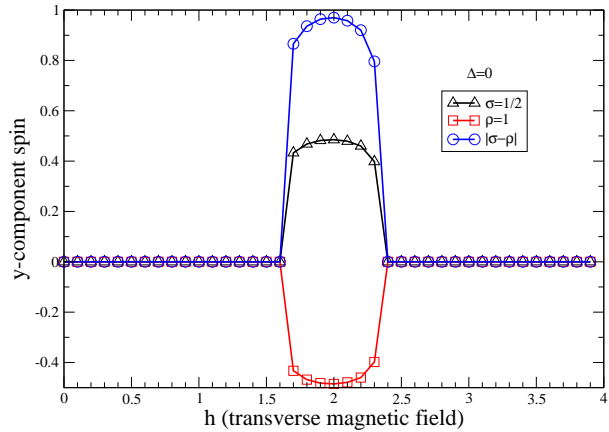}
\caption{ The $y$-component sublattice magnetization versus the transverse magnetic field
for $\Delta=0$. This component is nonzero only in the intermediate 
phase $1.6\lesssim h \lesssim  2.4$.
Moreover, the $y$-component spins are exactly equal and antiparallel for both sublattices.
The staggered magnetization of the unit cell in the $y$ direction is 
nonzero within the intermediate region.
}
\label{ymagnetization}
\end{center}
\end{figure}
To get more insight on the ground state properties of the model,
we have plotted the $y$-component spin expectation value versus the transverse
magnetic field in Fig. \ref{ymagnetization} for $\Delta=0$. The magnetization
in the $y$ direction for both sublattice spins is zero for $h\lesssim1.6$ and $h\gtrsim2.4$;
however, it becomes nonzero in the intermediate region $1.6\lesssim h \lesssim  2.4$.
The values of the $y$ component spins in the unit cell
are equal, and their directions are opposite to each other,
$\langle \sigma^y \rangle= - \langle \rho^y \rangle$. It is surprising that for any
value of the magnetic field $1.6\lesssim h \lesssim  2.4$ we get
$\langle \sigma^y \rangle= - \langle \rho^y \rangle$ whereas the spin magnitude
on the sublattices are different ($\sigma\neq\rho$).
At the factorizing field, $h_f\simeq 2.24$ (which will be explained
in the next section), where the condition 
$\sigma \sin |\theta| = -\rho \sin |\beta|$ should be satisfied, the mentioned 
relation is obtained $\langle \sigma^y \rangle= - \langle \rho^y \rangle$.
The staggered magnetization
in the $y$ direction, $SM_y=\langle \sigma^y - \rho^y\rangle$, is nonzero for this region.
Moreover, our numerical data verifies that the $z$ component magnetization on both
sublattices is zero for any value of the magnetic field.

Generally, let us consider the $y$ component staggered magnetization 
as an order parameter, which is nonzero for $h_{c1}(\Delta)<h<h_{c2}(\Delta)$ and zero
elsewhere. Nonzero $SM_y$ corresponds to a spontaneous breaking of Z$_2$ symmetry.
In fact, for any value of $\Delta$ the model has a Z$_2$ symmetry which can be expressed 
by the parity operator $P=\otimes_{i} \sigma_i^x \rho_i^x$. This symmetry, which
can also be considered as a $\pi$ rotation around the magnetic field direction ($x$),
leads to vanishing value for the $z$ and $y$ components of the spins. However, the symmetry
is spontaneously broken for $h_{c1}(\Delta)<h<h_{c2}(\Delta)$ which selects
one of the parity eigenkets to give nonzero sublattice magnetization in the $y$ direction.




\section{Spin wave analysis \label{swt}}

We have applied the spin wave theory 
to get more knowledge and a qualitative picture of the phase
diagram.
The SWT is a method to describe
a spin model in terms of boson operators. 
The elementary excitations of the spin model are given by bosonic quasi-particles
which are constructed on a given background. 
Based on this fact the SWT can
be considered on different backgrounds to build up a bosonic system.
Typically, a state in the Hamiltonian Hilbert space is considered as 
the background  which is supposed to be
the ground state within an approximation. However, there exists
some spin models such as the isotropic antiferromagnetic Heisenberg spin-1/2 chain 
that do not have
an ordered ground state, and thus the SWT fails to explain the properties of
the model correctly. Therefore, the existence of an exact ground state
is a good starting point to initiate a spin wave analysis.

The ferrimagnetic chain both in the absence and presence of longitudinal magnetic
fields has been studied by the SWT 
\cite{Pati 97, Yamamoto 99, Kolezhuk 99, Yamamoto 98, Sakai 99, Abolfath 01}.
Although the N\'{e}el state is not the exact ground state for a ferrimagnet in the
presence of a longitudinal field, the SWT gives a good description of the model which 
justifies that the
quantum fluctuations are not strong enough to ruin the whole picture. It would
be more interesting to initiate a spin wave theory based on an exact ground state
for a ferrimagnet in the presence of a transverse magnetic field.
According to Ref. \cite{Rezai 10}, the exact ground state of a general class of ferrimagnets
can be found at the factorizing magnetic field, $h=h_f$. This ground state is a factorized state,
which is a perfect background to implement SWT. It gives a reliable analysis 
around $h=h_f$ (see next subsection).
We will also study the SWT for small and large magnitudes of the 
magnetic field. Our analysis is limited to the linear spin wave approximation to 
get the magnetic properties of the spin-($\sigma, \rho$) ferrimagnets in the presence of
a transverse magnetic field.

\subsection{SWT at $h=h_f$ }

Let us briefly introduce the exact factorized ground state of a ferrimagnet in the 
presence of a magnetic field \cite{Rezai 10}. 
The factorized ground state for the Hamiltonian of Eq. (\ref{hamiltonian})
can be written in the following form:
\be
\label{fs}
|\psi_{0}\rangle=\bigotimes_{i\in A_{\sigma}, j\in B_{\rho}}|\sigma'_{i}\rangle
 |\rho''_{j}\rangle,
\ee
where $|\sigma'_{i}\rangle$ and $|\rho''_{j}\rangle$ are the eigenstates of
$\vec{\sigma}_i\cdot\hat{n}_i'$ and $\vec{\rho}_j \cdot\hat{n}_j''$
with the largest eigenvalues, respectively,
with $\hat{n}_i'$ and $\hat{n}_j''$ being
unit vectors pointing in polar angles ($\theta, \varphi=0$) and
($\beta, \alpha=0$). $A_{\sigma}$ and $B_{\rho}$ represent the
two sublattices which contain the two different spins.
The factorized state is called a {\it bi-angle state}, 
defined by the two angles ($\theta, \beta$) and represents the ground state of the model at $h=h_f$,
where 
\bea
\label{angles}
\cos \beta &=&\frac{\rho+\Delta\sigma}{\sqrt{\rho^2+\sigma^2+2\Delta\rho\sigma}}, \nonumber \\
\cos \theta &=&\frac{\sigma+\Delta\rho}{\sqrt{\rho^2+\sigma^2+2\Delta\rho\sigma}},
\eea
and 
\bea
h_{f}=2\sqrt{\rho^{2}+\sigma^2+2\Delta\sigma\rho},\nonumber \\
\epsilon_{f} = - (\sigma^2 +\rho^2 +\Delta \sigma\rho),
\eea
with $\epsilon_{f}$ being the ground state energy per site at the factorizing field.

To perform the spin wave analysis around $h=h_f$, we first implement a 
rotation on the original Hamiltonian ($H$).
The rotated Hamiltonian ($\tilde{H}$) is the result of
rotations on all lattice points of $H$, and is given by the
following relations
\bea
\tilde{H}&=& \tilde{D}^{\dag}H\tilde{D}, \nonumber \\
\tilde{D}&=&\bigotimes_{i\in A_{\sigma}, j\in B_{\rho}}
D_i^{\sigma}(0, \theta,0) D_j^{\rho}(0, \beta,0).
\eea
The rotation operator
\be
D^{\rho}(0,\beta,0)=D(\alpha=0,\beta,\gamma=0)=D_{z}(\alpha)D_{y}(\beta)D_{z}(\gamma),
\nonumber
\ee
is defined in terms of Euler angles, and a similar expression is considered for
$D^{\sigma}(0,\theta,0)$.

In the rotated basis  defined by ($x', y', z'$) and
($x'', y'', z''$), the bi-angle state becomes the 
fully polarized ground state 
 of $\tilde{H}$.
In the next step, the rotated Hamiltonian is bosonized via a
Holstein-Primakoff (HP) transformation, 
\begin{eqnarray}
\sigma^{+}_i=\sqrt{2\sigma-a^{\dag}_ia_i} \ a_i , \ \ \ \ \ \ \ \
\sigma^{x'}_i=\sigma-a_i^{\dag}a_i, \nonumber \\
\rho^{+}_j=\sqrt{2\rho-b^{\dag}_jb_j} \ b_j,         \ \ \ \ \ \ \ \
\rho^{x''}_j=\rho-b^{\dag}_jb_j,
\end{eqnarray}
where $a_i (a^{\dagger}_i)$ and $b_j (b^{\dagger}_j)$ are two types of annihilation (creation)
boson operators, satisfying the commutation relations: $[a_i, a_j^{\dagger}]=\delta_{i,j}$,
$[b_i, b_j^{\dagger}]=\delta_{i,j}$, $[a_i, b_j^{\dagger}]=0$ and $[a_i, b_j]=0$.

The Hamiltonian in the momentum ($k$) space and in the linear spin wave
approximation is written as
\begin{eqnarray}
\label{H}
&&\tilde{H}=E_{1}+\tilde{H}_{1}+\tilde{H}_{2},  \nonumber \\
&&E_{1}=N J \sigma\rho \cos(\beta-\theta)+N J h(\rho\cos\beta+\sigma\cos\theta), \nonumber \\
&&\tilde{H}_{1}=J \sum_{k} \biggl\{ 2 \sqrt{\rho\sigma} \Delta \cos\frac {k}{2}   ( a_{k}b_{k}^{\dag}+b_{k}a_{k}^{\dag}) \nonumber \\
 &&+\left(\frac{h^2_{f}}{2\rho}+\frac{h_f\sigma\cos\theta }{\rho}+(h_f-h)\cos\beta -2\Delta\sigma \right) b_{k}^{\dag}b_{k} \nonumber  \\
&&+\left( \frac{h^2_{f}}{2\sigma} + \frac{h_f\rho\cos\beta }{\sigma}+(h_f-h)\cos\theta-2\Delta\rho \right) a_{k}^{\dag}a_{k}\biggr\}, \nonumber \\
&&\tilde{H}_{2}=J \frac{\sqrt{N}(h_f-h)}{\sqrt{2}}\Big[\sqrt{\rho}\sin\beta\ (b_{0}+b_{0}^{\dag})
\nonumber \\
&& \;\;\;\;\;\;\; +\sqrt{\sigma}\sin\theta (a_{0}+a_{0}^{\dag})\Big],\nonumber \\
\end{eqnarray}
where $N$ is the total number of spins in each sublattice. 
The unitary transformation that diagonalizes $\tilde{H}_{1}$ is given by
\begin{eqnarray}
\chi_{k}&=&a_{k}\cos\eta_{k}- b_{k}\sin\eta _{k}, \nonumber\\
\psi_{k}&=& b_{k}\cos\eta_{k}+a_{k}\sin\eta_{k},
\end{eqnarray}
where $\chi_{k}$ and $\psi_{k}$ are the quasi-particle boson operators that
preserve the bosonic commutation relations.
In this representation, we obtain
\begin{equation}
\tilde{H}_{1}= \sum_{k}(\omega^{-}(k)\chi_{k}^{\dag}\chi_{k}+\omega^{+}(k)\psi_{k}^{\dag}\psi_{k}),
\end{equation}
in which
$\omega^{\pm}$ are the quasi-particle excitation modes. The dispersion relations are given by
\bea
\omega^{\pm}(k)&=& J D^+\pm J\frac{D^{-}+2\Delta\sqrt{\sigma \rho}\tan(2\eta_k)\cos\frac{k}{2}}{\sqrt{1+\tan^2(2\eta_k)}}, \nonumber \\
&&\tan(2\eta_{k})=\frac{2\Delta\sqrt{\rho\sigma}\cos\frac{k}{2}}{D^-},
\eea
in which
\bea
D^{\pm}&\equiv&\frac{h^2_{f}}{4}(\frac{1}{\rho}\pm\frac{1}{\sigma})+h_f(\frac{\sigma}{2\rho}\cos\theta\pm\frac{\rho}{2\sigma} \cos\beta )\nonumber \\
&-&\Delta(\sigma\pm\rho)+\frac{h_f-h}{2}(\cos\beta\pm\cos\theta).
\eea
A shift on the zero momentum component of boson operators, defined by
two constants $t^{\pm}$, diagonalizes the full Hamiltonian; i.e.,
$\chi_{0}\rightarrow \chi_{0}+t^{-} \;, \;
\psi_{0}\rightarrow  \psi_{0}+t^{+}$, where
\bea
t^{+}=\frac{\sqrt{2N} (h-h_f) \left(\sqrt{\sigma}\sin\eta_0\sin\theta+ \sqrt{\rho}\cos\eta_{0} \sin\beta\right)}{2\omega^{+}}, \nonumber \\
t^{-}=\frac{\sqrt{2N} (h-h_f) \left(\sqrt{\sigma}\cos\eta_0\sin\theta- \sqrt{\rho}\sin\eta_{0} \sin\beta\right)}{2\omega^{-}}.\nonumber \\
\label{tptm}
\eea
The diagonalized Hamiltonian is given by
\begin{eqnarray}
\label{dlh}
&&\tilde{H}=E_{gs}+\sum_{k}\bigg(\omega^{-}(k)\chi_{k}^{\dag}\chi_{k}+\omega^{+}(k)\psi_{k}^{\dag}\psi_{k}\bigg),
\nonumber \\
&&E_{gs}=E_1+ \omega^{-}t^{-^{2}}+ \omega^{+}t^{+^{2}} \nonumber \\
&&+\sqrt{2N} J (h_f-h)\bigg( (\sqrt{\sigma}\sin\theta\cos\eta_{0}-\sqrt{\rho}\sin\eta_{0}\sin\beta)t^{-}
 \nonumber \\
&&+(\sqrt{\sigma}\sin\theta\sin\eta_{0}+\sqrt{\rho}\cos\eta_{0}\sin\beta)t^{+}\bigg),\nonumber \\
&&E_f = 2N J \epsilon_{f}
\end{eqnarray}
where $E_{gs}$ is the ground state energy which reduces to $E_f$ at
the factorizing field ($h_f$) (i.e., the energy of the exact {\it bi-angle state}).

\begin{figure}
\begin{center}
\includegraphics[width=8cm]{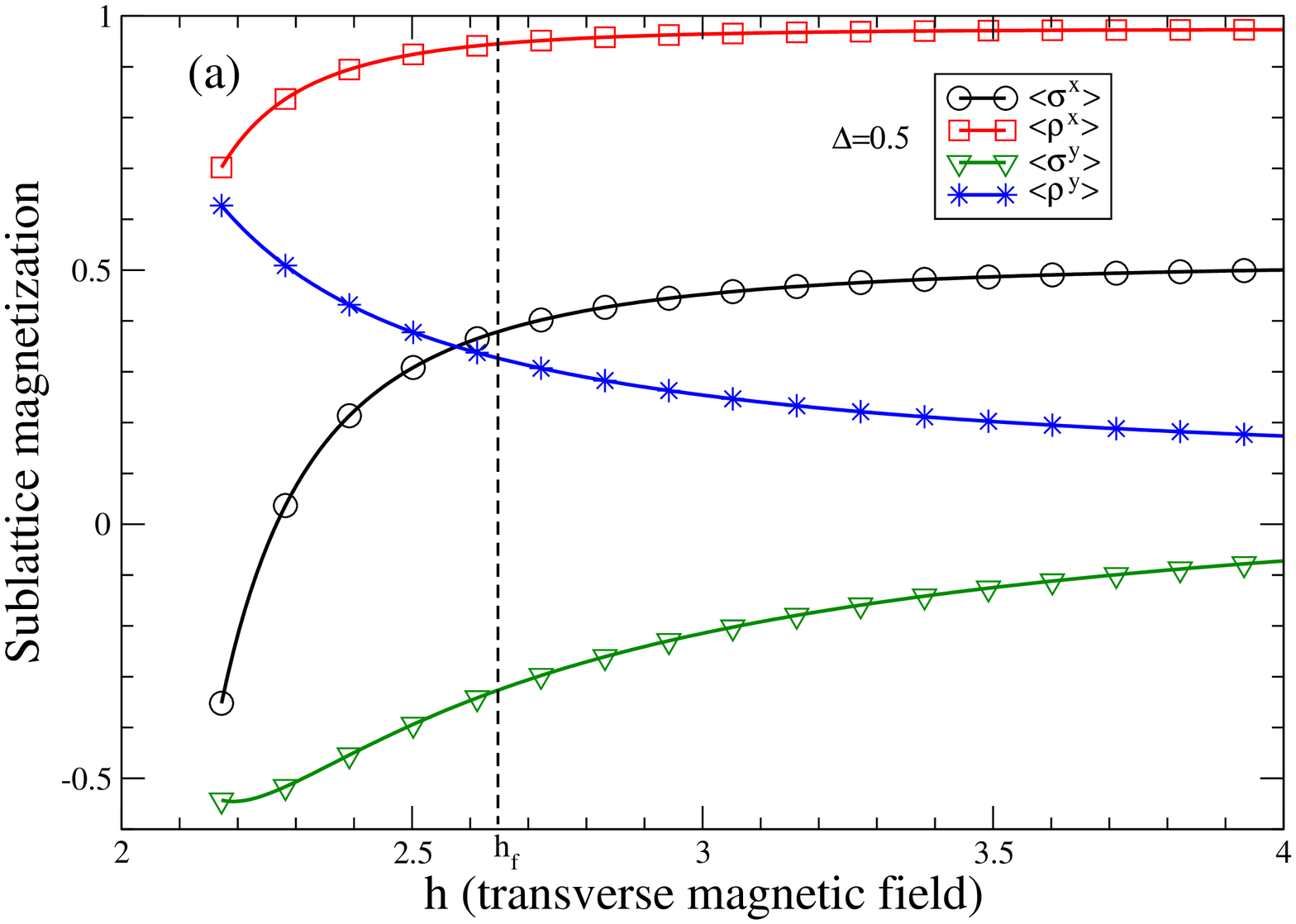}
\vspace{1cm}

\includegraphics[width=8cm]{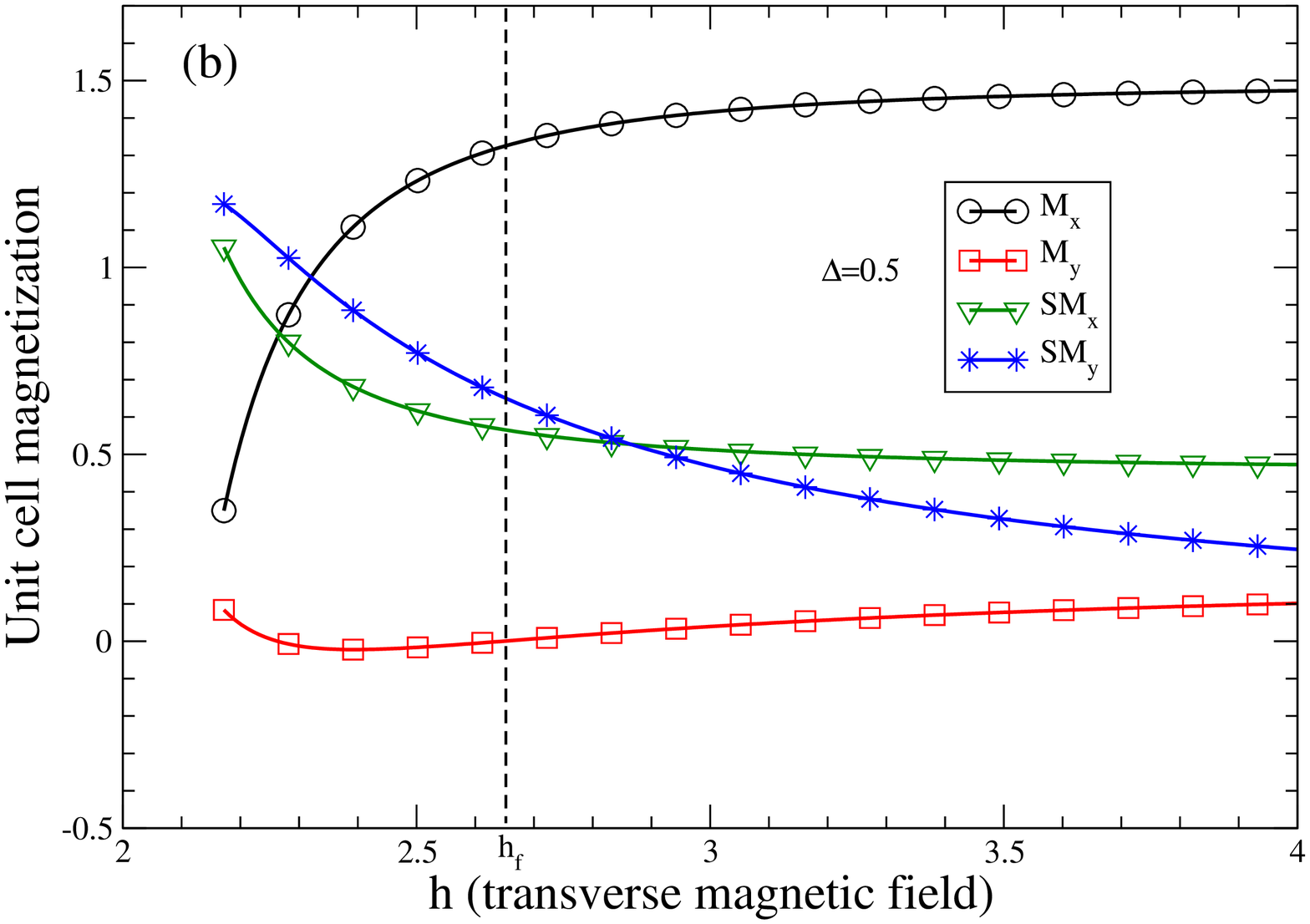}
\caption{(a) The sublattice magnetization. (b) Magnetization and staggered magnetization 
per unit cell of 
the anisotropic
ferrimagnetic spin-($1/2,1$) chain versus transverse field, for $\Delta=0.5$.
The factorized ground state is chosen as the 
background in the linear SWT.}
 \label{ffs}
\end{center}
\end{figure}

The magnetic properties of model (\ref{hamiltonian}) can be studied 
through  the linear spin wave theory-Eq. (\ref{dlh}).
In Fig. \ref{ffs}-(a), we have plotted the sublattice magnetization of 
the anisotropic ferrimagnetic spin-($1/2, 1$) chain for $\Delta=0.5$. 
The $x$ and $y$ components of sublattice magnetization are nonzero 
[Fig. \ref{ffs}-(a)]; however, the $z$-component of the sublattice magnetization is zero,
denoting that the spins are located in the $xy$ plane. It should be noted that 
the values of sublattice magnetization is exact at the factorizing 
field while it is approximately
correct for the magnetic field close to the factorizing field.
The $\alpha$-component of total magnetization per unit cell
is $M_{\alpha}=\langle \rho^{\alpha}+\sigma^{\alpha} \rangle$
and the corresponding staggered magnetization is defined 
$SM_{\alpha}=\langle \rho^{\alpha}-\sigma^{\alpha} \rangle$.
In Fig. \ref{ffs}-(b), we have plotted the $x$ and $y$ components of total 
magnetization
and staggered magnetization. 
Around the factorizing field the model has a considerable  magnetization 
in the $x$ direction and a staggered magnetization in the $y$ direction, 
which identifies a spin-flop phase around the factorizing field. The model 
has a dual character i.e. it behaves like a  ferromagnet
in the $x$ direction and like an antiferromagnet in the $y$ direction, 
it is the result of two branches
of excitations, Eq. (\ref{dlh}), which are the origin of the existence of 
two dynamics in the model \cite{Siah 08,Abou 10}. We will discuss later 
the effects of magnetic field on the configuration of both spins in more details.

\subsection{SWT at weak and strong magnetic fields}
(a) {\it Weak Field SWT} 

In the SWT it is  assumed that the ground state
defines a particular classical direction for the spins. In the weak magnetic fields
close to $h=0$, we expect to have a N\'{e}el-ordered configuration. Therefore,
we use the following Holstein-Primakoff (HP) transformations:
\begin{eqnarray}
\sigma^{+}_i=a_i^{\dagger} \sqrt{2\sigma-a^{\dag}_ia_i}  , \ \ \ \ \ \ \ \
\sigma^{x}_i=-\sigma+a_i^{\dag}a_i, \nonumber \\
\rho^{+}_j= \sqrt{2\rho-b^{\dag}_jb_j} \ b_j,         \ \ \ \ \ \ \ \
\rho^{x}_j=\rho-b^{\dag}_jb_j.
\end{eqnarray}
In the linear spin wave approximation and within Fourier space representation, 
one can diagonalize the Hamiltonian which is given by
\begin{equation}
H=E_0 +\sum_{k}\{\nu^{-}(k)v_{k}^{\dag} v_{k}+\nu^{+}(k) w_{k}^{\dag}w_{k}\},
\label{WFSWT}
\end{equation}
where
\bea
E_0&=& -  N J ( 2  \sigma \rho + \rho + \sigma ) - N J h ( \rho - \sigma ) 
+\frac{1}{2}\sum_{k}(\nu^{-}(k)+\nu^{+}(k) ),  \nonumber \\
&&\nu^{\pm}(k)= J \sqrt {2(p^2+s^2-2\Delta\rho\sigma\cos^2\frac{k}{2} \pm D_1)} , \nonumber \\
D_1&=&\sqrt {(p^2  - s^2 )^2  -4[\Delta(p^2  + s^2)-ps(1 + \Delta ^2 )]\rho \sigma \cos^2\frac{k}{2}} \nonumber\\
p&=&\rho -\frac{h}{2},\ \ \ \ \ \ \ \ \ \ \ \ \  s=\sigma + \frac{h}{2}.
\label{weakspectrum}
\eea
and $v^{\dagger}_k, w^{\dagger}_k (v_k, w_k)$ are bosonic quasi-particle creation
(annihilation) operators. 
The procedure of the diagonalization \cite{paraunitary} dictates that the bosonic Hamiltonian
should be positive definite. This constraint implies that
for $|\Delta |\leq 1$ the amount of magnetic field obeys the condition $h< 2(\rho -\sigma)$, and for  
$1\leq|\Delta|<\frac{\rho+\sigma}{2 \sqrt{\sigma \rho}}$
the magnetic field should be
$|h-\rho+\sigma| <\sqrt{(\rho+\sigma)^2-4\rho \sigma \Delta^2}.$
\begin{figure}
\begin{center}
\includegraphics[width=8cm]{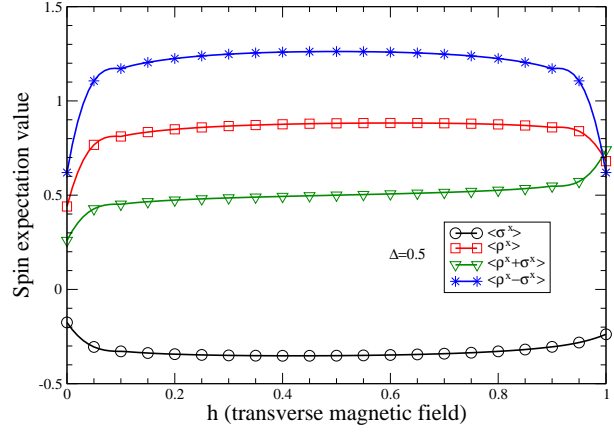}
\caption{The sublattice magnetization, total magnetization and staggered magnetization 
per unit cell of the anisotropic
ferrimagnetic spin-($1/2,1$) chain versus transverse field, for $\Delta=0.5$, when
the N\'{e}el order is chosen as the background in the linear SWT.}
 \label{neel}
\end{center}
\end{figure}

Let us consider the special case of ($\sigma=\frac{1}{2},\rho=1$) and $\Delta=0.5$. 
The Hamiltonian of this system (in the linear SWT approximation) is positive 
definite only for magnetic fields  smaller than $h_0=1$. 
Accepting this condition we have plotted in Fig. \ref{neel} 
the sublattices field-induced magnetization, 
the total magnetization and 
the staggered magnetization per cell of 
the whole chain versus transverse field $h$.
It shows that for $0<h<1$ the model is affected slightly by
the transverse magnetic field. In other
words, the staggered magnetization in the $x$ direction is close to its maximum value 
(the N\'{e}el ordered state).
The quantum fluctuations for $0<h<1$ are not strong enough to change the magnetization from
its zero field value. However, upon reaching $h_0=1$ the quantum fluctuations are suddenly
increased so that they destroy the ordered state completely. 
Thus within this linear SWT, the first
critical field is  $h^{SWT}_{c1}=h_0=1$ and for an arbitrary ($\sigma , \rho$)-ferrimagnetic 
chain it becomes $h^{SWT}_{c1}=2|\rho -\sigma|$.
Although $h^{SWT}_{c_1}$ does not depend on the anisotropy parameter $\Delta$
and is slightly different from the DMRG results (Fig. \ref{xmagnetization}),
the linear SWT describes the elementary excitations of the model well.

(b) {\it Strong Field SWT} 

For the strong magnetic fields the ground state is ordered in the direction
of the magnetic field. The fully polarized ground state in which all spins are aligned
in the field direction is used as the background for initiating the SWT.
In this case the following HP transformation is implemented for the spin operators
\begin{eqnarray}
\sigma^{+}_i=\sqrt{2\sigma-a^{\dag}_ia_i} \ a_i , \ \ \ \ \ \ \ \
\sigma^{x}_i=\sigma-a_i^{\dag}a_i, \nonumber \\
\rho^{+}_j=\ \sqrt{2\rho-b^{\dag}_jb_j} \ b_j ,         \ \ \ \ \ \ \ \
\rho^{x}_j=\rho-b^{\dag}_jb_j.
\end{eqnarray}
The diagonalized Hamiltonian in terms of the Fourier space representation and within the
linear SWT is
\begin{equation}
H=E_0 +\sum_{k}\{\Omega^{-}(k)V_{k}^{\dag} V_{k}+\Omega^{+}(k) W_{k}^{\dag}W_{k}\},
\end{equation}
where
\bea
E_0&=&N J( 2 \sigma \rho + \rho + \sigma ) - N J h ( \rho + \sigma + 1 ) 
+ \frac{1}{2}\sum_{k}(\Omega^{-}(k)+\Omega^{+}(k) ) , \nonumber \\
&&\Omega^{\pm}(k)= J \sqrt {2(p^2+s^2+2\Delta\rho\sigma\cos^2\frac{k}{2} \pm D_2)} , \nonumber \\
D_2&=&\sqrt {(p^2  - s^2 )^2  +4[\Delta(p^2  + s^2)+ps(1 + \Delta ^2 )]\rho \sigma \cos^2\frac{k}{2}} \nonumber\\
p&=&\frac{h}{2}-\rho,\ \ \ \ \ \ \ \ \ \ \ \ \  s=\frac{h}{2}-\sigma,
\label{strongspectrum}
\eea
and $V^{\dagger}_k, W^{\dagger}_k (V_k, W_k)$ are bosonic quasi-particle creation
(annihilation) operators.
The condition to have a positive definite bosonic Hamiltonian implies that
for $|\Delta|\leq 1$ the amount of the magnetic field should be larger than $2(\rho +\sigma)$
and for $|\Delta|\geq1$ the magnetic field should be larger than 
$\rho+\sigma+\sqrt{(\rho-\sigma)^2+4\rho \sigma \Delta^2}$.
\begin{figure}
\begin{center}
\includegraphics[width=8cm]{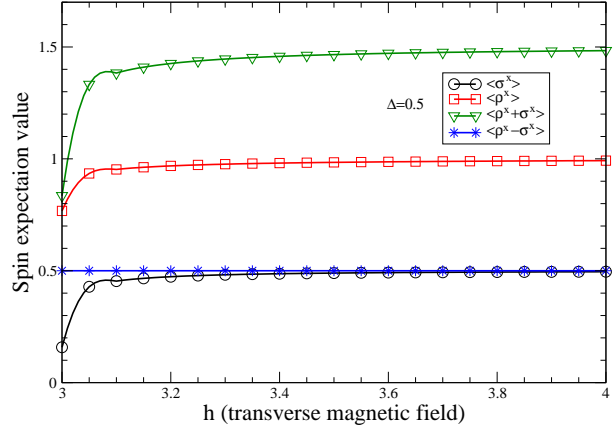}
\caption{The magnetization of sublattices, the total magnetization, 
and the staggered magnetization per unit cell of an anisotropic
ferrimagnetic spin-($1/2,1$) chain versus transverse field and for $\Delta=0.5$ 
and when
the background in the linear SWT is the field-induced fully polarized state.}
 \label{fero}
\end{center}
\end{figure}

Again we consider the special case of ($\sigma=\frac{1}{2}, \rho=1$) and $\Delta=0.5$. 
The Hamiltonian of this system in the linear SWT approximation is positive definite only for 
a magnetic field larger than $h^{SWT}_{c2}=3$. The magnetization of each sublattice, 
the total field-induced magnetization, and  the staggered magnetization per unit 
cell are plotted in Fig. \ref{fero}. For $h>3$, the model is in the polarized phase.
We have already shown in Ref. \cite{Rezai 10, Abou 10} that the full saturation only happens
for the isotropic case $\Delta=1$. Thus the model possesses an upper critical field
$h_{c2}=3$ for $\Delta=1$. The comparison with DMRG results shows that $h^{SWT}_{c2}=3$
is the true value, which is the consequence of weak quantum fluctuations for the strong
field regimes.
For $\Delta\neq1$, the fully saturated state appears
at infinite magnetic field. It can be understood simply by imposing
$\theta=0=\beta$ in Eq. (\ref{angles}) which can be fulfilled only for $\Delta=1$
in the Hamiltonian given by Eq. (\ref{hamiltonian}). In general, the full saturation occurs
at a  finite magnetic field if the model has the U(1) symmetry
around the direction of the magnetic field.

Let us discuss qualitatively the effects of a non commuting transverse magnetic field on 
the phase diagram of the anisotropic ferrimagnetic spin-($1/2, 1$) chain.
The SWT gives two branches of quasi-particle excitations for each of the
small, intermediate and large magnetic field regions. At zero magnetic field the lower branch
is gapless with ferromagnetic nature while the upper one is gapped with antiferromagnetic
signature. A nonzero magnetic field opens a gap in the ferromagnetic branch which remains
robust for $h\leq h_{c1}$. Moreover, the staggered magnetization in the field direction
is close to its maximum value which implies a N\'{e}el phase.
 At $h=h_{c1}$ a quantum phase transition from the N\'{e}el phase to the spin-flop phase
takes place where the staggered magnetization perpendicular to the field direction
becomes nonzero. The quasi-particle excitations for the spin-flop phase are given
by $\omega^{\pm}(k)$. In the spin-flop phase ($h_{c1}<h<h_{c2}$) 
an entanglement phase transition
occurs at $h=h_f$ where the quantum correlations become independent for
$h<h_f$ and $h>h_f$. The increase of magnetic field causes the second quantum 
phase transition at $h=h_{c2}$ to a nearly polarized state in the field direction.
The excitations in the field induced polarized phase ($h>h_{c2}$) are gapful given
by $\Omega^{\pm}(k)$, where the gap is proportional to the magnetic field.



\section{Summary and discussion\label{summary}}

The ground state phase diagram of the anisotropic ferrimagnetic ($\sigma, \rho$) chain 
in the presence of a non commuting transverse magnetic field has been studied.
The general picture has been obtained within the spin wave approximation.
We have applied three schemes of linear spin wave approximation to 
find the magnetic phase diagram of the anisotropic ferrimagnetic 
spin-($\sigma, \rho$) chain with anisotropy parameter $\Delta$ and in the presence of 
the transverse magnetic field ($h$).
The spin wave approximation has been applied close to $h=0$ (weak fields), $h=h_f$ (intermediate
regime), and $h\gg h_f$ (strong fields), where $h_f$ is the factorizing magnetic field.
The ground state is known exactly at $h=h_f$ as a product of single spin states. 
We have studied the magnetization in the field direction. There is a plateau
at $M_x=0.5$ for isotropic case where the ground state energy is linear in magnetic field while
no plateau observed for the anisotropic cases. 
However, the magnetization along the magnetic field 
changes slightly as long as $h\leq h_{c1}$ and its value is $M_x\simeq0.5$, 
which motivates to recognize it as a N\'{e}el phase . 
The model exhibits a quantum phase transition  at $h=h_{c1}$
from the N\'{e}el phase to (i) a spin-flop phase
for $\Delta\neq1$, (ii) a gapless Luttinger liquid for $\Delta=1$ \cite{Kolezhuk 99, Abolfath 01}. 
The magnetization evolves in the spin-flop phase when the magnetic
field is increased. The spin-flop phase contains
the factorizing field ($h=h_f$) where an entanglement phase transition takes place and
quantum correlations vanish. Further increase of the magnetic field leads 
to a polarized phase which resembles a plateau at the saturated magnetization 
in the field direction. However, it will be fully saturated only for $\Delta=1$ 
(the presence of a rotational symmetry around the magnetic field) which is 
represented by a quantum phase transition at a finite value $h_{c2}$.
The validity domain of spin wave analysis were introduced and 
it was shown that the corresponding results were
in good agreement with the DMRG numerical computations.

\begin{figure}[ht]
\vspace{1cm}
\centerline{\includegraphics[width=8cm,angle=0]{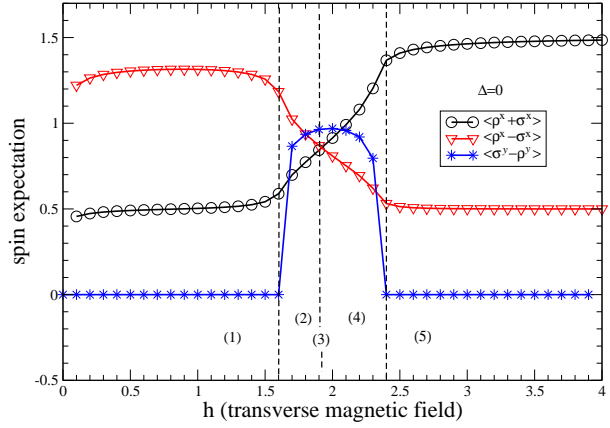}}
\caption{The $x$-component magnetization, $x$- and $y$-components staggered magnetization versus 
the transverse field for a ferrimagnetic spin-($1/2, 1$) chain. 
Effects of the magnetic field on the spins of each sublattice are divided into five different regions.
}
\label{XYmagnet}
\end{figure}

To get more accurate values on the magnetization process of spin-($1/2, 1$) ferrimagnet, 
we have also plotted in Fig. \ref{XYmagnet} the DMRG data of
the $x$- and $y$-component staggered magnetization in addition to the 
$x$-component magnetization of unit cell versus the transverse magnetic field for
$\Delta=0$. The magnetization
curve has been divided to five regions which has been labeled 
in fig. \ref{SFS}, fig. \ref{XYmagnet}, and also in Table. \ref{table1}.
Region-(1) is defined by the N\'{e}el phase for $0\leq h<h_{c1}\simeq1.6$ where
both $M_x$ and $SM_x$
are nearly constant  while $SM_y$
is zero. The spin-flop (gapped) phase, $h_{c1} \leq h \leq h_{c2}$, 
where a nonzero $SM_y$ sets up
can be distinguished to three parts, namely regions-(2-4). 
For $h_{c1}\leq h \lesssim 1.9$ which is labeled region-(2) we observe
$\langle\sigma^x\rangle<0$ and $\langle\rho^x\rangle>0$. It is a spin-flop phase
which is called spin-flop (I) in Table. \ref{table1}.
Region-(3) is defined at $h\simeq 1.9$ where the 
projection of smaller spin along the magnetic field becomes zero, 
$\langle\sigma^x\rangle=0$, i.e. $M_x=SM_x$. The rest, 
$1.9 \lesssim h \leq h_{c2}\simeq 2.4$, labeled by region-(4)
where $\langle\sigma^x\rangle>0$ and $\langle\rho^x\rangle>0$ is
called spin-flop (II). 
The region-(5) is the polarized phase
along the direction of magnetic field, i.e. $M_x\simeq 1.5$ and $SM_y=0$.
It is observed from Fig. \ref{xmagnetization}-(a) that the component of smaller spin 
in the direction of the magnetic field is
affected strongly by the magnetic field while the corresponding component 
for the larger one is almost constant.

The spin-flop (I) is a characteristic behavior
of XXZ {\it ferrimagnets} in the presence of transverse magnetic field 
because the spin component of the smaller spin along the
magnetic field is opposite to the field direction ($\langle\sigma^x\rangle<0$)
while the spin-flop (II) is similar to the corresponding phase of the
{\it homogeneous} XXZ spin chain in the presence of 
transverse magnetic field ($\langle\sigma^x\rangle>0$) \cite{Langari 04,Abou 10}. 
In the anisotropic ferrimagnetic chain the transverse field
first develops a N\'{e}el phase and a field-induced quantum phase transition
leads to a spin-flop phase.
Moreover, the Z$_2$ symmetry is spontaneously broken for small-field region
in the homogeneous spin chain while it will be broken in the intermediate fields
$h_{c1}(\Delta)<h<h_{c2}(\Delta)$
for ferrimagnets. A summary of different properties of the homogenous XXZ spin 1/2 chain
and the corresponding ($1/2, 1$) ferrimagnet both for isotropic and anisotropic cases
is presented in Table. \ref{table2}.


\begin{table}[ht]
\caption{Different configurations of the ground state of the 
ferrimagnetic spin-($1/2, 1$) chain with $\Delta=0$ in the presence of a transverse magnetic field.}
\footnotesize\rm
\label{table1}
\begin{tabular*}{\textwidth}{@{}l*{15}{@{\extracolsep{0pt plus12pt}}l}}
\hline
Region&$h$& Phase & Order parameters\\
 \hline
(1)&$0\leq h<1.6$& N\'{e}el &$M_x=1/2, SM_y=0$\\
(2)&$1.6\leq h<1.9$& Spin-Flop(I) &$\langle\sigma^x\rangle<0, SM_y>0$\\
(3)&$h\simeq1.9$ &  Spin-Flop &$\langle\sigma^x\rangle=0, SM_y>0$\\
(4)&$1.9\leq h<2.4$& Spin-Flop(II) &$\langle\sigma^x\rangle>0, SM_y>0$\\
(5)&$h>2.4$& Nearly Polarized &$M_x\simeq3/2, SM_y=0$\\
 \hline
\end{tabular*}
\end{table}

It is also interesting to mention that the low energy effective Hamiltonian of the
anisotropic spin-($1/2, 1$) chain in the presence of
a transverse magnetic field
can be represented by the fully anisotropic (XYZ) spin-1/2 Heisenberg chain in
an applied field (though we do not report such calculations in this paper). 
This helps to get more knowledge from the
results on the effective model \cite{0208216}. However, both spin wave approximation
and DMRG results show that the model has two nearly constant magnetization in the
presence of transverse magnetic field, the small-field plateau at $M_x\simeq 0.5$ 
for $h<h_{c1}(\Delta)$
and the saturated $M_x\simeq1.5$ for large fields ($h>h_{c2}(\Delta)$).
The general
behavior is the same for any value of the anisotropy parameter ($\Delta$); however,
the critical fields $h_{c1}(\Delta)$ and $h_{c2}(\Delta)$ depend on $\Delta$.
For instance, $h_{c1}(\Delta=0.5)\simeq1.8$ and $h_{c2}(\Delta=0.5)\simeq2.6$.

\begin{table}[ht]
\caption{Different ground state phases are classified for the 
heterogeneous spin-($1/2, 1$) XXZ ferrimagnet along with 
the homogeneous spin $1/2$ XXZ antiferromagnet.
The comparision between isotropic  and anisotropic cases
in the presence of the transverse magnetic field ($h$) is presented.
The magnetization per unit cell is $m$.
The ferrimagnet has two critical points $h_{c1}$ and $h_{c2}$ while the homogeneous
antiferromagnet has a critical point at $h_c$.}
\footnotesize\rm
\label{table2}
\begin{tabular*}{\textwidth}{@{}l*{15}{@{\extracolsep{0pt plus12pt}}l}}
 \hline
Spin& Region & Isotropic case ($\Delta=1$)& Anisotropic case ($\Delta\neq1$) \\
\hline
($1/2, 1$) &$0\leq h<h_{c1}$& Gapped N\'{e}el, plateau at $m=1/2$ &Gapped  N\'{e}el, no plateau\\
($1/2, 1$) &$h_{c1}< h<h_{c2}$& Gapless Luttinger liquid, no plateau & Gapped spin-flop, no plateau\\
($1/2, 1$) &$h > h_{c2}$ & Gapped paramagnet, plateau at $m=3/2$ & Gapped paramagnet, no plateau\\
 \hline
$1/2$ &$0\leq h< h_c $& Gapless spin-fluid, no plateau & Gapped spin-flop, no plateau\\
$1/2$ &$h>h_c$& Gapped paramagnet, plateau at $m=1/2$ & Gapped paramagnet, no plateau\\
 \hline
\end{tabular*}
\end{table}

The magnetization process can also be viewed as a {\it non-unitary evolution} of the system.
The entanglement of a pure state (ground state in our case) is conserved under
local unitary operations  \cite{Bennett 96}.
 For the ferrimagnetic spin-($1/2, 1$)  chain, the entanglement of the system is decreased 
by increasing the magnetic field for $h<h_f$. 
The entanglement vanishes at $h=h_f$ where the ground state is given by a 
tensor product state. This is an entanglement phase transition. It is thus concluded that
the effect of magnetic field is a non-unitary evolution of the ground state.

\section{acknowledgment}
J.A thanks H. Movahhedian for his fruitful comments.
A. L. would like to thank A. T. Rezakhani for his detailed comments 
on the final version of the manuscript.
A.L and M.R. would like to thank 
the hospitality of physics department of the institute for research
in fundamental sciences (IPM)  during part of this collaboration.
This work was supported in part by the Center of Excellence in
Complex Systems and Condensed Matter (www.cscm.ir).
The DMRG computation has been done by using ALPS package \cite{ALPS}
which is acknowledged.

\end{document}